# Selection of Alternative Local Interconnect Metals: Beyond Traditional Criteria Towards Sustainable and Secure Supply Chains

Corresponding Author: Lizzie Boakes[a]: lizzie.boakes@imec.be

Co-authors: Lars-Åke Ragnarsson[a]: lars-ake.ragnarsson@imec.be , Cédric Rolin[a]: cedric.rolin@imec.be, Christoph Adelmann[a]: christoph.adelmann@imec.be

[a]imec, Remisebosweg 1, 3001 Leuven, Belgium

## Abstract

In response to aggressive scaling demands in semiconductor manufacturing and the growing need to apply sustainable practices, this paper presents a holistic sustainability assessment framework for evaluating alternative metals for advanced applications. The framework, consisting of seven sustainability aspects, aims to guide researchers and industry stakeholders towards decisions fostering a more sustainable and secure future for microelectronics. This study applies the framework to assess the sustainability of alternative local interconnect metals. The framework identifies five metals (Ti, Al, Ni, Co, and Mo) with relatively favourable performance in at least six out of nine specific indicators, while others (Pt, Ru, Ir, Rh, and Pd) exhibit poorer sustainability metrics. The study recommends further analyses, suggesting the incorporation of case-specific functional units and the use of normalization and weighting factors for a comprehensive evaluation. Coupled with traditional technological assessments, this framework equips decision-makers with essential tools to broaden criteria for selecting alternative metals, aligning semiconductor manufacturing with broader sustainability objectives.

## Keywords

## 1. Introduction

Today the most aggressively scaled local interconnects in advanced IC technology nodes have reached dimensions on the order of 10 nm. This has driven conventional metals, such as copper and tungsten, to their physical limits. The pursuit of reduced electron scattering, better electromigration hardness, and continued scaling through barrier and liner layer removal has prompted the search for alternative metals to achieve improved reliability and lower line resistance (Adelmann et al., 2014; Gall et al., 2021; Rigsby et al., 2021; Tokei et al., 2016).

The decision criteria for selecting alternative metals for interconnects have traditionally centred around their intrinsic physical properties and behaviour in small dimensions, specifically resistivity magnitude and atomic migration (Moon et al., 2023). Additionally, the decision criteria extended to their manufacturability and associated costs (Baklanov et al., 2015; Baruah et al., 2023; Samsung, 2018). However, the necessity of including sustainability aspects (SAs) in the decision criteria has become increasingly apparent in recent years. For example, recent geopolitical events suggest that export restrictions, such as those imposed by China to regulate the export of gallium and germanium (Kabir, 2023), may play a non-trivial role in future international markets for raw materials (Kowalski and Legendre, 2023), impacting metal availability. Given that certain alternative metals are classified as 'strategic', it emphasizes the need to consider supply chain resilience to mitigate vulnerabilities arising from geopolitical tensions and export regulations.

In addition, ecological and social SAs are emerging as critical factors in the selection of alternative metals. As the global IC chip market is expected to grow with a compound annual growth rate of 7 % between 2021 and 2030 (Sperling, 2022), designing future technology nodes to minimize the cradle-to-gate environmental impact of upstream materials is imperative to minimize the total ecological footprint of the semiconductor industry. Moreover, ensuring ethical practices in raw material extraction and production is crucial to promoting social responsibility in the semiconductor industry.

This paper initially discusses the adoption and significance of a life cycle thinking (LCT) approach within the proposed framework. It defines the functional unit for sustainability analysis and discusses factors related to incorporating process integration considerations. It then dedicates one section to each SA as shown in Table 1. Finally, it demonstrates the framework using a set of current and alternative interconnect metals: Cu, W, Ti, Ta, Al, Ni, Pt, Ru, Co, Mo, Ir, Rh and Pd. The paper concludes with a summary of the findings and insights.

*Table 1: Sustainability aspects with corresponding indicator and unit considered in the proposed sustainability assessment framework.*

| Sustainability aspect (SA) | Proposed indicator | unit | Relation to UN SDGs (Department of Economic and Social Affairs, 2023) | Relation to life cycle phase illustrated in Figure 2 |
|---|---|---|---|---|
| Supply risk | Herfindahl-Hirschmann Index for country concentration (HHI), market price, and price rate of change (ROC) | -, USD, and % | 12 | 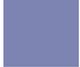 |
| Criticality and conflict | Presence on local Critical Raw Material (CRM) and Conflict Mineral lists | Yes/No | 12, 16 | 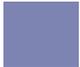 |
| Circularity | UL 3600 standard for site circularity of material flows | % | 12 | 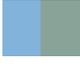 |
| Impact on climate change | LCIA – Global Warming Potential (GWP) | kg CO2 eq | 13 | 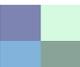 |
| Water use | LCIA – Water use | m³ world eq | 6 | 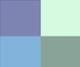 |

| Sustainability aspect (SA) | Proposed indicator | unit | Relation to UN SDGs (Department of Economic and Social Affairs, 2023) | Relation to life cycle phase illustrated in Figure 2 |
|---|---|---|---|---|
| Impact on natural resources | LCIA – Abiotic Depletion Potential (ADP) | kg Sb eq | 15 | 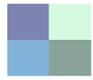 |
| Impact on human health | LCIA – Human toxicity Cancer and Non-Cancer, and Particulate Matter | Human Comparative Toxic Units (CTUh) and Disease incidence | 3 | 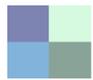 |

## 2. Methodology

To complete a comprehensive evaluation of alternative interconnect metals, we propose a set of decision criteria which differentiates technical and sustainability aspects, listing specific areas of concern (Figure 1). The seven SAs proposed are categorized into the three fundamental dimensions of sustainability: economic/governance, environmental, and social factors, following the Triple-Bottom-Line (TBL) framework (Ahmad et al., 2019). The seven SA were selected with the aim of covering diverse and mutually exclusive sustainability concerns identified in previous literature (Dewulf et al., 2015) as well as aligning the specific indicators with at least one of the UN Sustainable Development Goals (SDGs), shown in Table 1.

This paper presents a sustainability assessment framework for the seven SAs enclosed in the yellow box in Figure 1. The proposed framework is streamlined to facilitate eco-design practices and encourage process engineers to consider SAs while selecting alternative materials for advanced logic or memory interconnect applications during the R&D phases. The framework draws inspiration from previous research (Dewulf et al., 2015) that established an integrated sustainability assessment framework (ISAF) for the production and supply of raw materials and primary energy carriers. However, this current study focuses specifically on emerging local interconnect metals within the semiconductor industry, marking the first instance of such a specialized sustainability assessment framework.

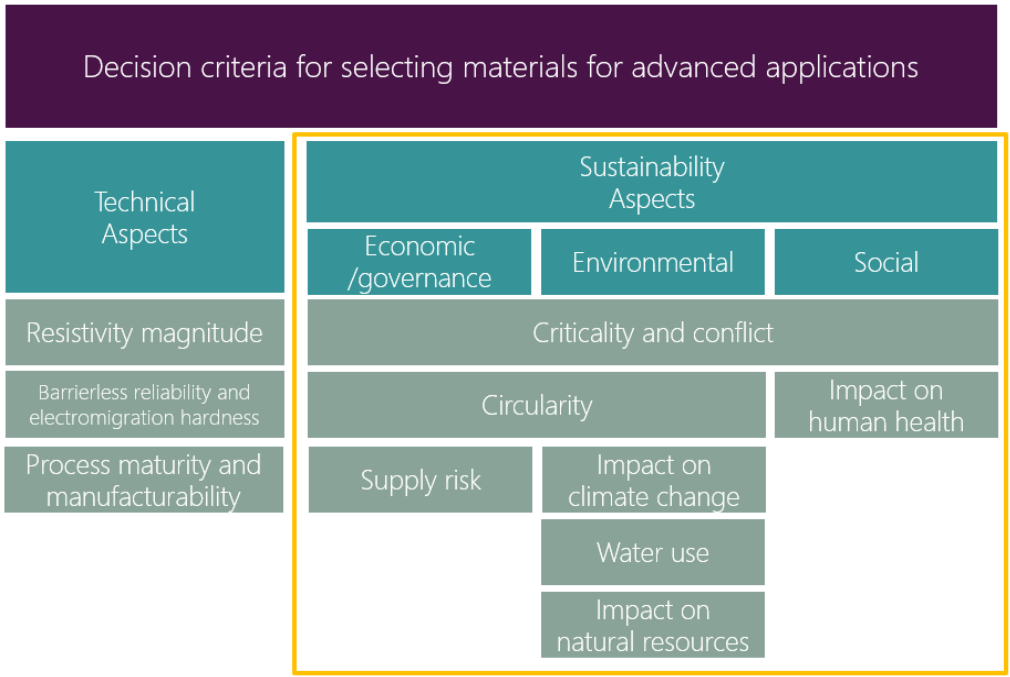

*Figure 1: Broadened decision criteria for selecting materials for advanced applications during the R&D phases. Technical and sustainability aspects are broken down into corresponding specific areas of concern.*

## 2.1. Life cycle thinking approach

LCT is described as a way of observing and reflecting, which leads to effective solutions for the overall improvement of the sustainability of products, processes, and systems (Mazzi, 2019). Adopting an LCT approach in this sustainability assessment framework provides a holistic system view, preventing the transfer of environmental burdens. Furthermore, it effectively identifies hotspot areas for improving sustainability. Therefore, it is important to contextualize the proposed framework within the life cycle of an interconnect metal, as illustrated in Figure 2.

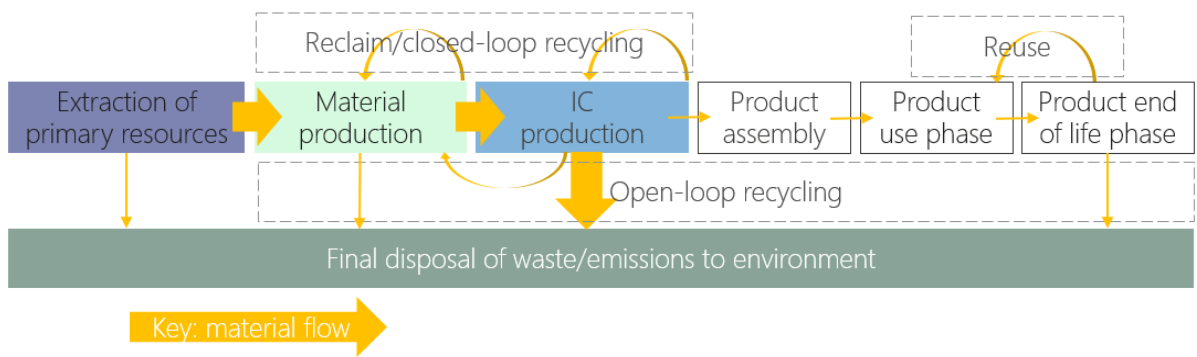

*Figure 2: Generic life cycle of an interconnect metal used in integrated circuit (IC) manufacturing. Each box represents a phase in the life cycle which consists of various processes, a gate signifies the transition between two life cycle phases. Each phase has a flow of interconnect metal (yellow arrows) and energy inputs and output of waste and emissions (not illustrated). The four most significant life cycle phases for sustainability analysis of interconnect materials are coloured.*

The material flow of the interconnect metal (yellow arrows in Figure 2) connects the life-cycle phases. All phases consume resources and generate waste, which, generally, should be reduced through measures such as enhanced process efficiency to improve sustainability. Only a very small fraction of the used materials ends up in the Si chip. Therefore, these four life cycle phases (coloured in Figure 2) have been highlighted as the focus of the sustainability assessment.

The functional unit of the interconnect metal and how it varies for the alternative metals are important considerations in the final sustainability assessment. The relative volumetric impacts (RVI) in Table 3 are calculated based on the embedded impact per volume of metal produced. This volume-based functional unit assumes that the volume of the final deposited layer of interconnect metal is independent of the metal, as the interconnect level of a specific technology node is defined by wire and via height, width, and length. RVI accounts for the differences in metal densities. However, it is important to note that different integration methods and surface topography influence the amount of metal required to reach the final deposited volume, these factors are dependent on metal choice and IC applications. Section 2.2 provides guidance on how to adapt the RVI values to a specific case study. Further guidance on defining a functional unit can be found in ISO 14040 (Lee and Inada, 2004).

In the context of metal production, the allocation of environmental burdens, with specific emphasis on metal companionability, represents a crucial consideration. Most alternative interconnect metals are "companion metals," denoting their status as by-products depending on the formation of deposits and their preferred host minerals (Eilu and Törmänen, 2021). Within the metals industry, economic allocation is commonly used to distribute environmental impact proportionately among valuable outputs (Nuss and Eckelman, 2014). This distribution is carried out by assigning a share of the total environmental impact, relative to the market price of each product output, to individual valuable outputs. Allocation is one of the most difficult components of a sustainability assessment and care should be taken to apply appropriate methods, particularly for highly companionable product systems, further information can be found in the ISO 14040 standard (Lee and Inada, 2004).

## 2.2. Integration considerations

In the pursuit of an LCT approach, wherein the displacement of environmental burdens is avoided, it is important to consider the integration method governing the deployment of alternative interconnect metals. Understanding material and energy flows surrounding the integration process is pivotal for understanding the overall sustainability of alternative interconnect metals. Generally, process flows using fewer process steps have a smaller environmental impact, but the energy requirements of the tools used are also important to consider if the data is available.

The electrical requirements on an interconnect metal go beyond resistivity and include metrics like electromigration hardness, and barrierless reliability. Therefore, assuming a fixed volume for the benchmarking of the different metal options is an oversimplification. The volume ratio to achieve the same function should be considered. First, the efficiency of the metal deposition methods must be evaluated. This is essential to establish the actual used volume ($V_{used}$) of the metal needed to achieve the desired volume of deposited metal. The efficiency of typical deposition processes ($Eff_{deposition}$) is in the range of 1 to 20% (Weber et al., 2023). Furthermore, the integration of the interconnects should be considered. Typical integration schemes are subtractive, meaning that after materials are deposited, they are later removed using etching. This results in further material loss, determined by the material use efficiency of the integration process ($Eff_{integration}$), and on the interconnect metal choice as well. Subtractive schemes will typically have an $Eff_{integration}$ of less than 50%. $V_{used}$ is defined in equation (1).

$$V_{used} = V_{deposited} / (Eff_{deposition} * Eff_{integration}) \quad (1)$$

The results in Table 3 provide the RVI, which is the ratio of the environmental impact to produce 1 cm³ of an interconnect metal with respect to the environmental impact for the production of 1 cm³ of Cu ($EI_{Cu}$), (provided in Table 4). The environmental impact of the interconnect metal is defined in equation (2).

$$EI_{metal} = V_{used} * RVI_{metal\ to\ Cu} * EI_{Cu} \quad (2)$$

## 3. Sustainability aspects

### 3.1. SA1: Supply risk

Supply chain risk for advanced interconnect metals can be approximately quantified using the Herfindahl–Hirschman index (HHI), which is a commonly accepted measure of market concentration (Eurostat, 2021). The main benefit of the HHI is its simplicity: it is calculated by squaring the global market share of each firm competing in the market and then summing the resulting numbers, as shown in equation (3).

$$HHI = \sum_{i=1}^{n} S_i^2 \quad (3)$$

$$s = percent\ market\ share\ of\ each\ firm\ (i)$$

A market can be classified into three categories based on the HHI value: Below 1,500 is considered a competitive marketplace, an HHI of 1,500 to 2,500 is moderately concentrated, and an HHI of 2,500 or greater is highly concentrated (Bromberg, 2023). A competitive market is a good indicator of low supply risk and thus a sustainable supply chain. The HHI value for the interconnect metals in Table 2, have been categorised based on this classification and highlighted green, amber, and red, respectively. Cu has the lowest HHI value of 1097, whereas Ru, Ir, and Rh have HHI values of 8718, 7986, and 7352, respectively, representing an extremely concentrated market with higher supply risk. It is important to recognise the inherent limitations of the HHI value that stem from its simplicity. Publicly available sources of HHI data often don't provide market-specific data, e.g., semiconductor grade metals, but instead provide a value for generic metal production, potentially misrepresenting the situation for that specific segment of the market. HHI values in Table 2 for the primary production of metals were taken from the annual World Mining Data report published by the Austrian Federal Ministry of Finance (World Mining Congress, 2023), HHI values for Ru and Ir were taken from the RMIS – Raw Materials Information System published by the European Commission's Joint Research Centre (Joint Research Centre, n.d.) as they were not available in the former source. HHI values may exhibit slight variations depending on the data source, so it is recommended to compare values from the same source if feasible.

Market price and price fluctuation, e.g., Price Rate of Change (ROC), are two other indicators to consider when evaluating supply risk. Metals with a high average price per kg pose higher risks when the economic climate worsens. Table 2 shows the average annual price for 2021 in USD (not specific to semiconductor-grade metals) (National Minerals Information Center, 2023). This data shows a large variation in the price (greater than 200,000 times the difference between the lowest and highest price). This indicates which interconnect metal is economically sustainable. Note that market price data for semiconductor-grade metals would be more relevant and should be prioritized if available. The ROC is defined in equation (4).

$$ROC = \left(\frac{Closing\ price_p - Closing\ price_{p-n}}{Closing\ price_{p-n}}\right) \times 100\% \quad (4)$$

Where *Closing price $_p$* is the closing price for the most recent period. *Closing price $_{p-n}$* is the closing price *n* periods prior (Mitchell, 2023).

### 3.2. SA2: Criticality and Conflict

Materials are classified as critical if their availability is important to national security or economic growth (Ashby, 2022). Governments have lists of Critical Raw Material (CRM), by definition, these criticality lists depend on the country's requirements and priorities. For example, Table 2 shows slight variations in the United States and European lists, for example, Cu is considered critical in the EU and not in the US. Amongst the advanced interconnect metal candidates, only Mo is absent in both the US and the EU CRM lists (Commission et al., 2023; Fortier et al., 2022). The CRM that should be considered in the sustainability analysis, is that of the country/region where the material is being consumed.

Being listed as a critical raw material can influence market dynamics, causing fluctuations in pricing, demand, and investment. Moreover, it can attract attention from policymakers, potentially resulting in the development of new policies and regulations which can hinder global material markets. Whilst using a listed CRM as an advanced interconnect material is not necessarily unsustainable, recognizing a CRM in your supply chain highlights the need for proactive measures to ensure a secure and sustainable supply, as well as triggering regulatory attention and requirements related to responsible sourcing, reporting, or recycling.

Table 2 also shows whether the metal is present on the EU conflict minerals list, which indicates which minerals are associated with politically unstable areas, where armed groups use forced labour to mine minerals and use the profits to fund their activities. Out of the four conflict minerals defined by the EU (Commission E, 2023), tungsten and tantalum are both considered alternatives for advanced interconnect metals. Avoiding the use of conflict minerals improves the social sustainability of the supply chain.

### 3.3. SA3: Circularity

To quantify circularity, one must measure the current or potential reuse/reclaim and recycling techniques applied to a specific scope. For interconnect metals, the scope which should be considered encapsulates the IC manufacturing processes, as this is where the majority of the material flow is consumed, and the principal metal waste flow is created. A circularity assessment of interconnect metals requires an in-depth understanding of the integration process used during IC production and the treatment processes applied to the waste streams that are generated. This is explained further in this section.

The mass balance of the interconnect metal is calculated for the process steps that consume it. The chemical form and origin/destination of the input/output flows are used to calculate the site material circularity index (*CI*), following the UL 3600 standard (ANSI/UL Standard for Safety, 2023). Figure 3 illustrates exemplar circularity assessment boundaries for an IC manufacturing process area. In both open and closed-loop material treatment processes there is material loss which should be accounted for, i.e., only the usable material flow is considered as a circular output of the IC process. The recycled input must be semiconductor-grade metal (the quality of the interconnect metal cannot be compromised). The *CI* is calculated using the equations (5-7) which reference to the flows in Figure 3.

$$CI_{upstream} = (Closedloop_{input} + Recycled_{input})/Total_{inputs} \quad (5)$$

$$CI_{downstream} = (Closedloop_{output} + Recycled_{output})/Total_{outputs} \quad (6)$$

$$CI = (CI_{upstream} + CI_{downstream})/2 \quad (7)$$

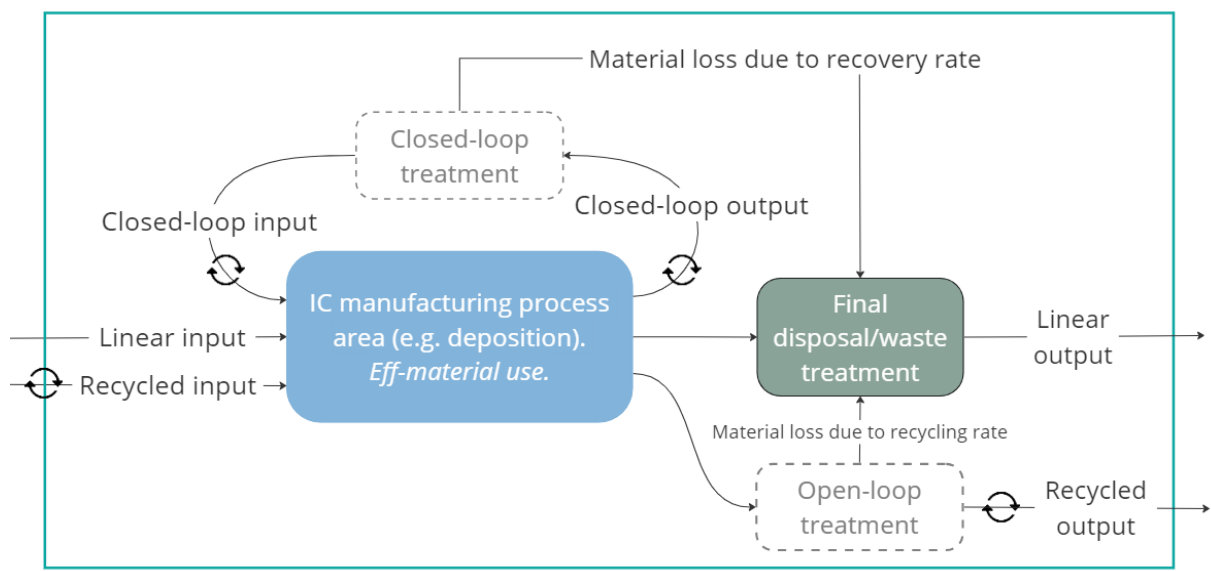

*Figure 3: Example circularity assessment boundaries (green box) applied to an IC manufacturing process area. All input and output flows are measured per functional unit. The final volume of interconnect metal remaining on the wafer after all processing steps (e.g., deposition and CMP) is assumed to be insignificant.*

Avoiding the production of waste streams is prioritized over increasing the circularity index, following the waste hierarchy. Therefore, material use efficiency ($Eff_{material\ use}$) within the process step is important to consider. $Eff_{material\ use}$ is defined in equation (8). An example of sustainable practice would be to increase $Eff_{material\ use}$, e.g., improving $Eff_{deposition}$ during PVD or CVD, which typically are in the range of 5-10 % for interconnect metals (meaning that 90-95% of the input material goes to the waste stream) (Weber et al., 2023).

$$Eff_{material\ use} = 1 - \frac{Total_{outputs}}{Total_{inputs}} = \frac{Final\ deposited\ mass}{Input\ mass} = 1 - \frac{Wasted\ mass}{Input\ mass} = Eff_{deposition} * Eff_{integration} \quad (8)$$

A qualitative evaluation of the cost and availability of separation processes for waste streams containing metals, which increases the recycling potential, serves as a good initial step in comparing the circularity potential of alternative metals. The highest priority goes to recovery solutions capable of recycling interconnect metals within a closed-loop system (e.g., onsite reclamation processes). In this system, the metal is purified to semiconductor-grade and substitutes a portion of the original virgin metal input. Alternatively, recycling can occur in an open-loop system where the purity of the metal has been compromised, allowing it to be recycled in another sector. It is important to note that all waste processing systems consume energy and materials, creating a trade-off between the additional environmental costs associated with the waste processing and the savings achieved by substituting virgin material. This trade-off is not captured in the circularity rate calculation as per the UL 3600 standard. This trade-off can be captured by applying LCA principles to the recovery system, like the proposed LCA matrix model in (Schwarz et al., 2021).

Scientific literature, and industry-relevant papers/white papers are possible resources to assess the circularity potential. For example, an article by Chi and Tseng describes the extensive work that TSMC has been doing to recycle liquid copper waste, where TSMC has demonstrated a successful closed-loop system wherein semiconductor-grade copper is returned to the fabrication process for further use (Chi and Tseng, 2018). These sources allow for a qualitative understanding of the level of maturity and accessibility of existing recycling methods, as well as emerging recycling prospects for novel materials.

### 3.4. SA4: Impact on climate change

Global Warming Potential (GWP) is an indicator that characterizes the environmental impact, ($EI_{GWP}$), caused by the greenhouse gases (GHG) that are emitted into the atmosphere and are the root cause of a significant shift in global temperatures (Intergovernmental Panel On Climate Change, 2023). Figure 4 illustrates the ideal assessment boundaries to quantify the upstream, direct, and downstream GHG emissions for the IC production process involving the interconnect metal. If data availability prohibits conducting an assessment with these boundaries, a simplified assessment of the environmental impact of GWP ($EI_{GWP}$) that focuses solely on the upstream emissions, also known as embedded impact, stemming from the production of input interconnect metal consumed during the IC manufacturing process can be used.

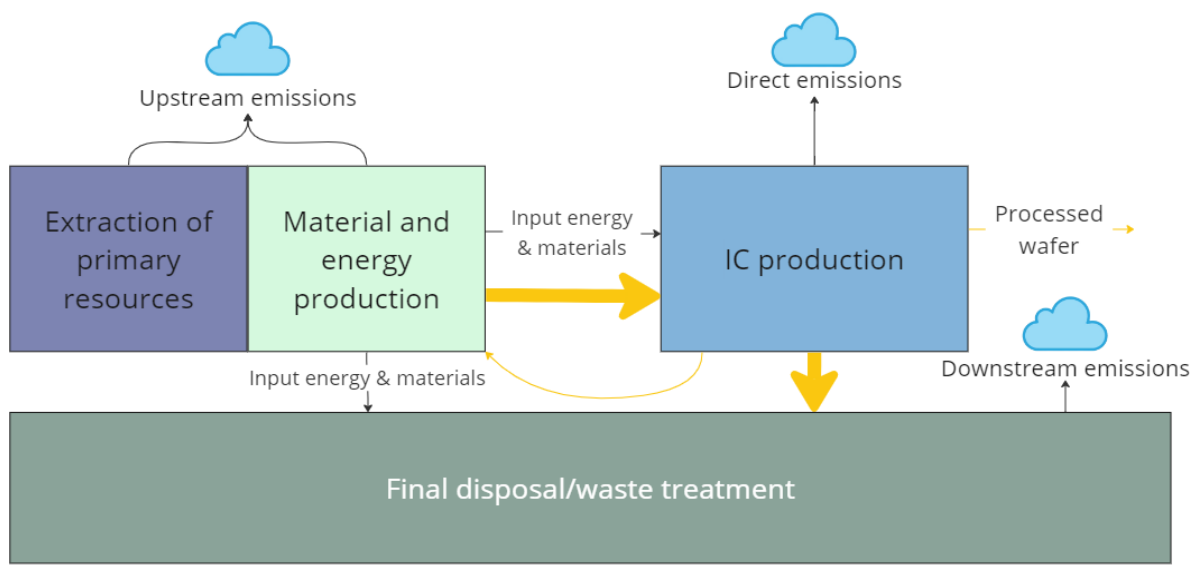

*Figure 4: Ideal assessment boundaries including upstream emissions resulting from the cradle-to-gate production of input electricity and materials, direct emissions from the IC production process, and downstream emissions resulting from the treatment of output material flows. The yellow arrows represent the interconnect metal flow.*

This simplified $EI_{GWP}$ is the product of the embedded impact to produce 1 cm³ of a given interconnect metal ($GWP_{i\ volume}$), the total volume used ($V_{used}$), as shown in equation (9). $GWP_{i\ volume}$ is a product of the embedded impact to product 1 kg of metal ($GWP_{i\ mass}$) and the metal density ($\rho_i$), as shown in equation (10). $V_{used}$ is defined in equation (1), see section 2.2 for further guidance.

$$EI_{GWP} = GWP_{i\ volume} * V_{used} \quad (9)$$

$$GWP_{i\ volume} = GWP_{i\ mass} * \rho_i \quad (10)$$

The relative embedded *GWP* values provided in Table 3 are the ratio between the $GWP_{i\ volume}$ of the alternative metal and $GWP_{Cu\ volume}$. $GWP_{i\ mass}$ values were taken from a life cycle assessment (LCA) study of metal production processes (Nuss and Eckelman, 2014), and multiplied by their respective metal densities (Semicore Inc, n.d.). It should be noted that these GWP values are likely to be underestimates of semiconductor-grade metals which have extremely high purity requirements. However, they can be used to do an initial relative comparison between the alternative metals.

High EI$_{GWP}$ values, resulting from the production of Rh, Pd, Ir, Ru and Pt, are caused by several factors such as complex and energy-intensive production procedures which rely on fossil-based fuels as an energy source. For example, Cu open-pit or underground mining, milling, and refining processes are energy-intensive but have been optimized to require less energy and emit fewer greenhouse gases compared to Rh production which involves more complex and energy-intensive processes due to its scarcity and occurrence as by-product of Platinum Group Metals (PGM)s. Table 3 shows the GWP of Rh is 17363 times larger than Cu, indicating a significantly larger environmental impact to deposit the same volume of interconnect metal.

## 3.5. SA5: Water scarcity

Water use occurs in all four of the highlighted life cycle phases illustrated in Figure 5. Ideally, a full water balance for the processes within these four phases would be done to fully quantify the water scarcity impact. However, as this is a cumbersome task and data availability is often limited, this framework proposed a simplified system boundary of analysis, illustrated by the blue box in Figure 5. The proposed boundary incorporates the upstream water used to produce the interconnect metal.

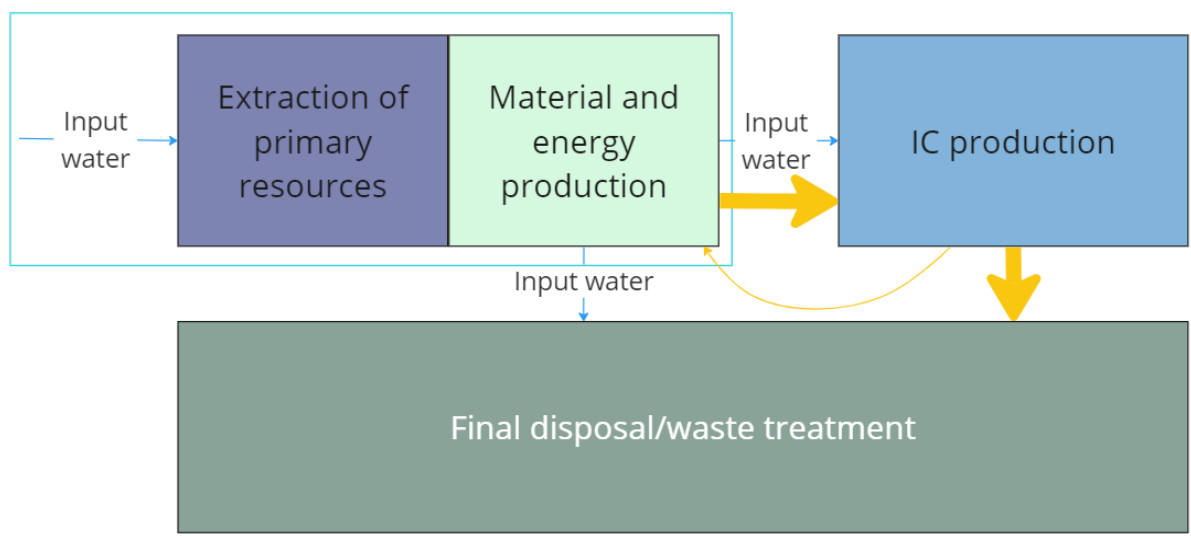

*Figure 5: Simplified assessment boundaries (blue box) surrounding the water used in the upstream cradle-to-gate processes for interconnect metal production. Yellow arrows represent the flow of the interconnect metal.*

Water scarcity (*WS*) has been quantified using proxy cradle-to-gate LCA processes for each interconnect metal (see Table 5). Proxy processes for Ru and Ir could not be found in available LCA databases, therefore their water scarcity values were calculated by multiplying the cradle-to-gate blue water consumption values (IPA, 2023), by the WULCA characterization factors (CF)s for an "unknown" geographic location (WULCA, 2019). 'Cradle-to-gate' in this context refers to all material and energy flows surrounding the life cycle processes: extraction of raw materials and material production, as illustrated in Figure 2. The 'water scarcity' values that are displayed in Table 3 are calculated using the EF 3.1 methodology, which adopts the Available WAter REmaining (AWARE) CFs developed by WULCA (Boulay et al., 2021).

The *WS* value to produce an interconnect metal, is the sum of the product of the country-specific $CF_{AWARE}$ and the volume of all blue water (non-agricultural) consumed, for a given country ($V_c$) for the total amount of countries (*n*) involved in the value chain (Fazio et al., 2018). The *WS* value is defined following equation (11). The EF 3.1 *WS* values per kg of interconnect metal production were multiplied by the interconnect metal density, resulting in a WS value in m³ world equivalent/cm³ of interconnect metal and set relative to Cu as shown in Table 3.

$$WS = \sum_{c=1}^{n} CF_{AWARE} \times V_c \quad (11)$$

This unit for WS offers the ability to incorporate geographical water scarcity variation when comparing cradle-to-gate water use during the production processes for the different interconnect metals, rather than simply considering the total volume of water consumed (Boulay et al., 2018). Ti, Al, and Ni result in a WS value of 2.68, 0.34, and 1.09 respectively, relative to Cu. The WS values of Pt, Ru, Ir, Rh, and Pd are all significantly larger in relation to Cu, with the largest relative value from Pt production, being 9819 times that of Cu.

## 3.6. SA6: Impact on natural resources

The depletion of abiotic resources is an important indicator to consider when comparing alternative metals for advanced interconnects. Abiotic resource Depletion Potential (ADP) is a method of impact assessment that has been widely adopted in life cycle assessment.

The characterization factor of ADP for a given resource/material *i* is defined by (Guinée and Heijungs, 1995) using equation (12).

$$ADP_i = {P_i R_{ref}^2} / {R_i^2 P_{ref}} \quad (12)$$

Here, $P_i$ is the world's annual production (kg/year) of a resource *i*, and $R_i$ is its ultimate reserve (kg). The $ADP_i$ is the ratio of resource *i* divided by the $ADP_{ref}$ of a reference resource, which usually is antimony (Sb).

A life cycle approach which considers all the mass and energy flows in the four highlighted life cycle phases in Figure 2 should ideally be used. Alternatively, a simplified ADP value can be quantified using a cradle-to-gate assessment for the production of 1 kg of interconnect metal, the impact values are expressed in kg Sb equivalent (van Oers et al., 2020). The values in Table 3 are calculated by multiplying the simplified ADP value by the metal density and setting it relative to Cu. Ti, Al, Ni, and Co have the lowest relative ADP values, while the largest ADP values are from Pt, Ru, Ir, and Pd. Pt has the largest ADP value which is 88,666 times larger than Cu.

### 3.7. SA7: Impact on human health

The impact on human health can be estimated using several indicators. Those proposed in this paper are from the EF 3.1 LCA methodology, namely, 'Human Toxicity Cancer and non-Cancer', and 'Particulate Matter'. These indicators were highlighted in a recent survey of LCA experts as the most worrisome impact indicators for human health (Commission et al., 2018).

Human toxicity (cancer and non-cancer) is measured in the Comparative Toxic Unit for human (CTUh), which represents the estimated increase in morbidity in the total human population. The calculation is based on USEtox® 2.1, which is a model that describes chemical fate, exposure, effect and optionally severity of emissions (USETOx®, n.d.). The values in Table 3 are calculated using the impact value per kg of metal production (Nuss and Eckelman, 2014), multiplied by the metal density, and set relative to Cu.

Particulate matter, also considered as the air pollution footprint, predicts the potential effect of fine dust emissions on human health. Fine particulate matter is mainly absorbed through the respiratory system, where it contaminates the lung alveoli and the bloodstream, promoting numerous illnesses (Thangavel et al., 2022). The indicator, expressed in the unit disease incidences, is calculated by applying the average slope between the Emission Response Function (ERF) working point and the theoretical minimum-risk level following the model developed by (Fantke P., 2016). The values in Table 3 are calculated using the impact values from the proxy LCA processes multiplied by the density of metal and set relative to Cu. Proxy processes for Ru and Ir could not be found in available LCA databases.

Table 3 shows that Mo results in the lowest impact value for particulate matter with only 0.39 Disease incidences relative to Cu. Pt, Rh, and Pd result in the highest impact values, with Pt resulting in the largest impact value: 27,685 larger than Cu.

Other SAs, for example, social SAs, could also be considered. A recent review paper by Hogrefe & Bohnet-Joschko (Hogrefe and Bohnet-Joschko, 2023), provides a thorough overview of corporate social sustainability research and provides examples of common themes and indicators used to access social sustainability. More specifically, Popovic et al. proposed a set of 31 quantitative social sustainability indicators which evaluate the supply chain (Popovic et al., 2018).

## 4. Discussion and conclusions

The pursuit of alternative local interconnect metals, driven by aggressive scaling demands, necessitates a comprehensive evaluation beyond traditional criteria. As geopolitical dynamics impact the availability of strategic metals, ensuring supply chain resilience and economic sustainability becomes indispensable. Concurrently, adopting environmentally and socially sustainable practices in material sourcing and manufacturing is vital for the microelectronics industry's responsible growth.

This paper describes a holistic sustainability assessment framework, which can steer researchers and industry stakeholders to make decisions for a more sustainable and secure future for semiconductor manufacturing. The framework comprises seven SAs, with at least one specific indicator for each. The indicators correspond to a specific phase of the life cycle of the interconnect metal used in IC processing. Although there are limitations to these indicators, mainly due to the availability of representative data, time, or LCA expertise, they provide a foundation for comparing relative SAs of the alternative interconnect metals. This framework applies an LCT approach and encourages the assessor to consider the material and energy flows surrounding the integration process to build a complete system view and avoid any displacement of environmental burdens.

Ti, Al, Ni, Co, and Mo performed relatively well in at least six out of the nine indicators whereas Pt, Ru, Ir, Rh, and Pd performed relatively poorly in at least five of the nine indicators in relation to the baseline (Cu). This initial assessment can already provide a first-hand approximation of which interconnect metals have a relatively large impact on the environment, and human health, or are generally a supply risk or economically unsustainable. This framework is the starting point for further analysis which would incorporate a case-specific functional unit to normalise the impact values. Furthermore, the impact values from some indicators, namely, GWP, WS, ADP, human toxicity – cancer, and particulate matter, could be multiplied by a weighting factor to normalize the impact values to a common unit, allowing relative impact between the different indicators to be compared. Normalization/weighting factors for the EF 3.1 LCIA methodology can be found in a recent JRC technical report (Andreasi Bassi et al., 2023). Alternatively, a case-specific weighting of the proposed indicators can help the assessor make decisions based on the impact results. For example, due to corporate sustainability commitments, one assessor may prioritize price stability/low supply risk and low impact on human health over criticality or ADP, which will restrict the analysis to those interconnect metals that perform well in those impact categories.

This paper lays the groundwork for a comprehensive framework for evaluating sustainability, which could be coupled with a traditional technological assessment, providing decision-makers with tools to broaden the criteria for selecting alternative metals for advanced interconnects.

*Table 2: Sustainability aspect indicators: HHI, ROC, annual average price, and presence of the metal on criticality or conflict lists. The values for 'Price, ROC' and 'Price, annual average' are classified as green if they are equal to or less than the second quartile (Q2), amber if the value lies between Q2 and the third quartile (Q3), and red if the value is equal to or above Q3.*

| Interconnect metal | Density (Semicore Inc, n.d.) | SA1: Herfindahl–Hirschman index (World Mining Congress, 2023) | SA1: Price, ROC (2018-2021) (National Minerals Information Center, 2023) | SA1: Price, annual average (2021) (National Minerals Information Center, 2023) | SA2: EU CRM (2023)/USGS Criticality list (2021)/EU Conflict List |
|---|---|---|---|---|---|
| | [kg/cm³] | [0-10000] | [%] | [USD/kg] | |
| Cu | 0.0090 | 1,097 | 43% | 9 | Yes/No/No |

| Interconnect metal | Density (Semicore Inc, n.d.) | SA1: Herfindahl–Hirschman index (World Mining Congress, 2023) | SA1: Price, ROC (2018-2021) (National Minerals Information Center, 2023) | SA1: Price, annual average (2021) (National Minerals Information Center, 2023) | SA2: EU CRM (2023)/USGS Criticality list (2021)/EU Conflict List |
|---|---|---|---|---|---|
| | [kg/cm³] | [0-10000] | [%] | [USD/kg] | |
| W | 0.0194 | 6203 | -14% | 28 | Yes/Yes/Yes |
| Ti | 0.0045 | 1598 | 0% | 11 | Yes/Yes/No |
| Ta | 0.0167 | 1658 | -26% | 158 | Yes/Yes/Yes |
| Al | 0.0027 | 3372 | 21% | 3 | Yes/Yes/No |
| Ni | 0.0089 | 2,110 | 41% | 20 | No/Yes/No |
| Pt | 0.0215 | 5690 | 24% | 35,183 | Yes/Yes/No |
| Ru | 0.0124 | 8,718 | 136% | 18,523 | Yes/Yes/No |
| Co | 0.0089 | 4,876 | -30% | 51 | Yes/Yes/No |
| Mo | 0.0102 | 2,266 | 31% | 35 | No/No/No |
| Ir | 0.0224 | 7986 | 299% | 165,846 | Yes/Yes/No |
| Rh | 0.0124 | 7352 | 810% | 651,184 | Yes/Yes/No |
| Pd | 0.0120 | 3250 | 133% | 77,778 | Yes/Yes/No |

*Table 3: Relative volumetric impact (RVI) values for interconnect metals with respect to Cu. SA5 and SA7 were quantified using EF 3.1 Water Use and EF 3.1 Particulate Matter, midpoint values using the cut-off allocation method. The proxy LCA processes are*

listed in Table 5. Values for each SA are classified as green if they are equal to or less Q2, amber if the value lies between Q2 and Q3, and red if the value is equal to or above Q3.

| Interconnect metal | SA4: Embedded GWP (Nuss and Eckelman, 2014) | SA5: Water scarcity (WS) (IPA, 2023; WULCA, 2019) | SA6: ADP (van Oers et al., 2020) | SA7: Human Toxicity (Cancer and non-cancer) (Nuss and Eckelman, 2014) | SA7: Particulate matter |
|---|---|---|---|---|---|
| Cu | 1.00 | 1.00 | 1.00 | 1.00 | 1.00 |
| W  | 9.72 | 12.07 | 2.24 | 0.27 | 30.72 |
| Ti | 1.47 | 2.68 | 0.00 | 0.01 | 4.16 |
| Ta | 173 | 60.36 | 0.11 | 0.83 | 97.86 |
| Al | 0.88 | 0.34 | 0.00 | 0.01 | 1.34 |
| Ni | 2.31 | 1.09 | 0.04 | 0.08 | 16.17 |
| Pt | 10687 | 9819 | 88666 | 816 | 27686 |
| Ru | 1040 | 7007 | 13806 | 82.81 | - |
| Co | 2.94 | 14.38 | 0.02 | 0.01 | 6.10 |
| Mo | 2.32 | 0.71 | 9.29 | 3.80 | 0.39 |
| Ir | 7911 | 8532 | 12963 | 463 | - |
| Rh | 17363 | 6477 | 0.11 | 1385 | 9297 |
| Pd | 1859 | 3770 | 48692 | 89.43 | 26121 |

Table 4: Environmental impact values to produce 1 cm³ of Cu using the LCA process 'GLO: Copper mix (99,999% from electrolysis)' from the Sphera LCA database.

| Sustainability aspect indicator | Unit | Cu |
|---|---|---|
| SA4: Embedded GWP | [kg CO2eq/cm³] | 2.51E-02 |
| SA5: Water scarcity (WS) | [m³ world equiv/cm³] | 2.35E-02 |
| SA6: ADP | [kg Sb eq/cm³] | 2.42E-04 |

| | | | |
|---|---|---|---|
| SA7: Human Toxicity (Cancer and non-cancer) | [CTUh/cm$^3$] | | 2.42E-06 |
| SA7: Particulate matter | [Disease incidences/cm$^3$] | | 5.11E-09 |

*Table 5: LCA processes used as proxies to model the production of semiconductor-grade interconnect metals. These LCA processes were used to quantify SA5 and SA7.*

| Interconnect metal | LCA process name | Data source | Process GUID |
|---|---|---|---|
| Cu | GLO: Copper mix (99,999% from electrolysis) | Sphera | 301D375B-4F27-43F2-BBE0-89F87CAE0DF1 |
| W | RoW: tungsten carbide powder production | Ecoinvent 3.9.1 | 81DC426A-310B-4DDD-BC12-CFD1E4D1D01B |
| Mo | RoW: molybdenum production | Ecoinvent 3.9.1 | E5B41016-A271-47BF-AA50-BE568C27A0F7 |
| Ti | GLO: titanium production | Ecoinvent 3.9.1 | C0AD6D5D-D118-44D4-BD44-A12786B49F78 |
| Ta | RoW: tantalum powder production, capacitor-grade | Ecoinvent 3.9.1 | 1E5C1F74-6760-4291-8327-B883371BB013 |
| Al | GLO: Aluminium ingot mix IAI 2015 | International Aluminium Institute (IAI) | 241D1242-4D0F-4DED-9A96-5181615B0BFB |
| Ni | GLO: Nickel (Class 1, >99.8% Nickel) | Nickel Institute | 04DC7156-8FDA-4C67-923E-E779ABD20E49 |
| Pt | GLO: Platinum, primary route | IPA | 2FFBD7FA-CBD3-4B70-A637-5F3133E30ED1 |
| Ru | - | - | - |
| Co | GLO: Cobalt, refined (metal) | Cobalt Institute (CI) | 935F46F9-1BD3-4412-84B8-09416221A0E3 |
| Mo | RoW: molybdenum production | Ecoinvent 3.9.1 | 4A726FF7-AFC3-4D56-84D7-0D248DF62581 |
| Ir | - | - | - |
| Rh | GLO: Rhodium, primary route | International Platinum group metals Association (IPA) | BDEFCB95-BB4C-4449-A28F-ED93D85FB428 |

| Pd | GLO: Palladium, primary route | IPA | EEBABC84-3436-4E5B-91D0-AFDF8CA2AFCC |